# Comparative Study of the Structural, Mechanical, Electronic, Optical and Thermodynamic Properties of Superconducting Disilicide YT$_2$Si$_2$ (X=Co, Ni, Ru, Rh, Pd, Ir) by DFT Simulation

Md. Atikur Rahman[a]*, Mahbub Hasan[a], Rukaia Khatun[a], Jannatul Ferdous Lubna[a], Sushmita Sarker[a], Md. Zahid Hasan, Wakil Hasan

DFT simulation based ab-initio approach has been executed for investigating the comparative study of the physical properties of superconducting disilicide materials YT$_2$Si$_2$ (T= Co, Ni, Ru, Rh, Pd, Ir). This is the first comparative theoretical investigation of these materials, which is done through Cambridge Serial Total Energy Package module. Extremely good relation has been observed between the synthesized and calculated structural parameters of all the superconductors. Mechanical structural stability of all the phases has been confirmed from the investigated elastic stiffness parameters. The investigated elastic moduli show good agreement with previously calculated data where available. Also the fundamental polycrystalline features such as bulk modulus, shear modulus, Young's modulus, Pugh's and Poison's ratios, ductile/brittle nature and hardness of superconducting YT$_2$Si$_2$ (T= Co, Ni, Ru, Rh, Pd, Ir) have been examined. Ductile nature of YRh$_2$Si$_2$, YPd$_2$Si$_2$ and brittle nature of YCo$_2$Si$_2$, YNi$_2$Si$_2$, and YIr$_2$Si$_2$ have been observed from analyzing of polycrystalline elastic parameters where YRu$_2$Si$_2$ lies on the border line of ductile/brittle nature. The high bulk modulus, Young's modulus, and hardness of YIr$_2$Si$_2$ ensured that this phase has high ability to resist volume and plastic deformation and suitable in industrial applications. On the other hand the material the small Young's modulus of YPd$_2$Si$_2$ ensured its application as a thermal barrier coating (TBC) material. Metallic nature of YT$_2$Si$_2$ (T= Co, Ni, Ru, Rh, Pd, Ir) has been confirmed from the band structure and DOS calculations. Mulliken atomic populations and charge density map reveal the existing of covalent and ionic bond in YT$_2$Si$_2$ (T= Co, Ni, Ru, Rh, Pd, Ir). Different optical features included dielectric function, refractive index, optical conductivity, reflectivity, absorption and loss function of YT$_2$Si$_2$ (T= Co, Ni, Ru, Rh, Pd, Ir) have been executed through CASTEP code directly. The high reflectivity in the UV energy region of these phases ensured their application as good solar reflector in this energy site. Furthermore the thermodynamic properties of superconducting YT$_2$Si$_2$ (T= Co, Ni, Ru, Rh, Pd, Ir) have been determined from the elastic stiffness constants. The high conductivity of YCo$_2$Si$_2$ is ensured from its high Debye and melting temperature. The minimum thermal conductivity of YPd$_2$Si$_2$ ensured that it is suitable to use as a thermal barrier coating (TBC) material.

**Keywords:** DFT simulation; Superconducting Phases YT$_2$Si$_2$ (T= Co, Ni, Ru, Rh, Pd, Ir); Structural and Mechanical Properties; Electronic and Bonding Properties; Optical and Thermal Properties

**Corresponding author:** atik0707phy@gmail.com

## 1. Introduction

Superconducting materials are always demandable for their valuable applications in modern technology and the discovery of high $T_c$ superconductors of several types of compounds has concerned huge attention in the recent time. The layered ternary compounds of 122 types belong to most inspiring and have been significantly studied by several researchers in several times. The 122 type ternary compounds have two types of structures; where one belong to body-centered tetragonal configuration $ThCr_2Si_2$ (space group I4/mmm) and the second one goes to primitive tetragonal structure $CaBe_2Ge_2$ (space group P4/nmm) [1]. There are more than seven hundred compounds fall in the $ThCr_2Si_2$-type configuration with space group of I4/mmm (139) [2-4]. At ambient condition the structure $ThCr_2Si_2$ is thermodynamically stable; on the other hand the structure $CaBe_2Ge_2$ is thermodynamically stable at high temperature. The general formula of $ThCr_2Si_2$-type configuration is $AT_2X_2$ where $A$ stand for alkali, alkaline earth or rare earth elements; $T$ stand for 3d, 4d and 5d transition metals and finally X belongs to p-elements such as Si, Ge, As or P. The crystal structure of $ThCr_2Si_2$-type was first synthesized and described by Ban and Sikirica in 1965 [5] however the theoretical simulation of $ThCr_2Si_2$ was first done by Shein and Ivanovskii in 2011 [6]. In 1996 Just and Pauer had give an entire geometric investigation of about 600 compounds of $ThCr_2Si_2$-type configuration [4]. From past to present most researchers have been working on $ThCr_2Si_2$-type compounds frequently for the reason that these types of compounds have several diversity of remarkable physical observable fact such as massive fermion character, superconductive manner (in high and low temperature), interesting magnetic order, assorted valency and valence instability, phase transition, Kondo effect etc [7-11]. The discovery of iron-based high $T_c$ (up to 49 K) superconducting materials $AFe_2As_2$ (A =Ca, Ba, Sr, Eu) fall in the $ThCr_2Si_2$-type configurations play a great role in the modern technology [12-14]. But it is so difficult to synthesis iron-based high $T_c$ superconductors in practically. To overcome this situation researcher was looking to discover the alternative of iron- pnictides and have succeeded by finding the mixture of ternary intermetallic $ThCr_2Si_2$-type compounds. The ideal $ThCr_2Si_2$-type crystal structure is obtained from several layers created by edge-sharing network of $BX_4$ tetrahedral alternating with Ba ions [4]. Several iron-free compounds included $CaPd_2Ge_2$ ($T_c$ = 1.98 K) [15], $CaPd_2As_2$ ($T_c$ = 1.27 K) [16], $SrPd_2Ge_2$ ($T_c$ = 3.04 K) [17], $SrIr_2Ge_2$, $SrNi_2Ge_2$ ($T_c$ < 1.8 K) [18], $YIr_2Si_2$ ($T_c$ = 2.52 K) [19], $BaNi_2P_2$ ($T_c$ = 2.80 K), $BaIr_2P_2$ ($T_c$ = 1.97 K), $BaRh_2P_2$ ($T_c$ = 0.70 K) [20], $YPt_2Si_2$ [21], $YPd_2Si_2$ and $YRh_2Si_2$ [22] show superconductive features with very low $T_c$ (below 3K). Very recent Pt, Ni, and Pd-based $ThCr_2Si_2$ type boro-carbides [23-25] show superconductive phenomena with transition temperature up to 23K. In this research article we are focused to comparative investigation of the details physical features included mechanical, hardness, anisotropic manner, properties related to photon and temperature of superconducting disilicide $YT_2Si_2$ (X=Co, Ni, Ru, Rh, Pd, Ir) along with some similar types of compounds for the initial moment. The studied compounds $YT_2Si_2$ (T= Co, Ni, Ru, Rh, Pd, Ir) belong to ternary layered $ThCr_2Si_2$-type tetragonal configurations where the set of lattice points fall in the space group I4/mmm (139) [21]. The compound $YCo_2Si_2$ was initially examined by Rossi *et al*. in 1978 via the X-ray powder diffraction method [26].

Höting et al. synthesized the silicides YT$_2$Si$_2$ (T = Co, Ni, Cu, Ru, Rh, Pd) by Solid-state NMR Spectroscopy in 2014 [27] where all the compounds go to ThCr$_2$Si$_2$-type structure. After reviewing the literature we have observed that only the theoretical study is performed on YIr$_2$Si$_2$ by I.R. Shein in 2011 [28] among all the studied compounds. No theoretical works on YT$_2$Si$_2$ (T= Co, Ni, Ru, Rh, Pd) are available in the literature yet. However a lot of theoretical works on the ThCr$_2$Si$_2$-type compounds are performed by several researchers in several times. M. U. Salma *et al.* studied the physical features of ARh$_2$Ge$_2$ (A = Ca, Sr, Y and Ba) in 2019 [29] through DFT simulation. In 2019 Chowdhury *et al.* investigated the details physical features of ThCr$_2$Si$_2$-type Ru-based materials SrRu$_2$X$_2$ (X= P, Ge, As) through ab-initio method [30]. Rahaman *et al.* studied the physical properties of Ir-based superconductors BaIr$_2$Mi$_2$ (Mi= P and As) [31] and Ru-based superconductors LaRu$_2$M$_2$ (M= P and As) [32] in 2017 using *ab*-initio method. In our previous study we have investigated the physical properties of Ni-based superconductors BaNi$_2$T$_2$ (T= P, As) [33] and SrNi$_2$M$_2$ (M= As and Ge) [34] in 2019. In 2019 Ali *et al.* discover the physical features of Pd-based superconductors XPd$_2$Ge$_2$ (X= Ca, Sr, La, Nd) [35]. Though remarkable efforts have been done to synthesis and discover the superconducting features of disilicides YT$_2$Si$_2$ (T= Co, Ni, Ru, Rh, Pd, Ir), there is still missing of detail theoretical works included electronic, mechanical, thermal and optical properties these superconductors in literature. On the basis of having important crystal features and practical applications of compounds YT$_2$Si$_2$ (T= Co, Ni, Ru, Rh, Pd, Ir) we therefore decided to study the all the physical attributes of these compounds employing the DFT simulation based on CASTEP code. Not only that we have tried to compare the investigated results with some similar type of compounds which are already available in literature. We think that this study will play an important role in the research area and superconductor based modern technology.

## 2. Computational Details

Employing Density Functional Theory (DFT) [36] we have examined the structural features, stability factors named elastic stiffness parameters, hardness, band structures, light and temperature related properties of superconducting disilicides YT$_2$Si$_2$ (T=Co, Ni, Ru, Rh, Pd, Ir) for first time. All the features are executed by CASTEP code [37]. The generalized gradient approximation (GGA) was used to estimate the exchange correlation functional energy accompanied with the Perdew-Burke-Ernzerhof functional (PBE) [38]. The Vanderbilt-type ultrasoft pseudopotential [39] was employed to describe the interactions happened between electrons and ions. We have sampled the Brillouin-zone using Monkhorst-Pack grids [40] of 14×14×6 with a plane wave cut-off energy of 500 eV for YCo$_2$Si$_2$ and YIr$_2$Si$_2$ and 16×16×7 with a plane wave cut-off energy of 520 eV for YT$_2$Si$_2$ (T= Ni, Ru, Rh, Pd) respectively. For searching the ground state of all the phases geometry optimization has been performed utilizing BFGS (Broyden-Fletcher-Goldfarb-Shanno) relaxing technique [41]. For the geometry optimization the convergence tolerance was fixed to ultra-fine quality along with energy of 5.0×10$^{-6}$ eV/atom, maximum force of 0.01eV/Å, maximum stress of 0.02 GPa and maximum displacement of 5.0×10$^{-4}$ Å. At the optimized structures the elastic stiffness constants are investigated using the finite stress-strain method [42] according to Hock's law. The 2D and 3D anisotropy contours are investigated through the ELATE code [43]. Photon related optical features of superconducting $YT_2Si_2$ (T=Co, Ni, Ru, Rh, Pd, Ir) have been studied by employing the CASTEP code [42] directly. The temperature dependent thermodynamic properties are calculated by utilizing the elastic stiffness constants.

## 3. Results and discussion

### 3.1 Structural Properties

The superconducting yttrium silicides $YT_2Si_2$ (T= Co, Ni, Ru, Rh, Pd, Ir) belong to ternary layered $ThCr_2Si_2$-type tetragonal configurations through space group of I4/mmm (139) [27]. Utilizing the synthesized lattice parameters we have drawn the conventional crystal structure of $YT_2Si_2$ (T= Co, Ni, Ru, Rh, Pd, Ir) which is displayed in Fig. 1 (a). Fig. 1(b) show the crystal structure of $YT_2Si_2$ (T= Co, Ni, Ru, Rh, Pd, Ir) which is attained after the geometry optimization. The unit cell of all the crystal structures contains two formula units with 10 atoms. In the unit cell of $YT_2Si_2$ (T= Co, Ni, Ru, Rh, Pd, Ir) the Y atom takes the position 2a (0 0 0), Co/Ni/Ru/Rh/Pd/Ir atoms are in the sites 4d (0 0.5 0.25) and at last the Si atom located at the sites 4e (0 0 $z_{si}$) where z represents the internal parameter. The primitive cell has one formula unit containing 5 atoms. The synthesized unit cell dimensions, atomic positions, Wyckoff positions, lattice parameters after optimization and cell volumes are listed in Table 1. The values of the investigated structural parameters are very close to the synthesized data confirming the accuracy of the present simulations.

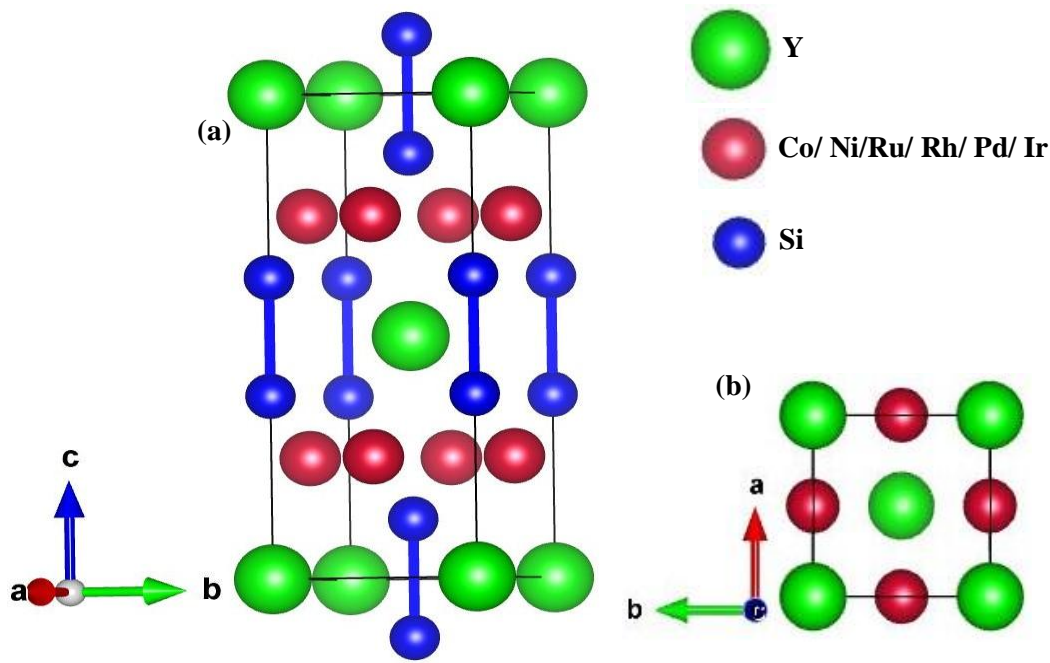

**Fig.5.1. (a)** Conventional 3D, and **(b)** the optimized 2D crystal structures of $YT_2Si_2$ (T= Co, Ni, Ru, Rh, Pd, Ir).

**Table 1:** Optimized lattice constants $a$, $c$ (in Å) and cell volume $V$ (in Å$^3$) of superconducting YT$_2$Si$_2$ (T= Co, Ni, Ru, Rh, Pd, Ir) polymorphs.

| Compounds | $a = b$ (Å) | $c$ (Å) | $V = a^2c$ (Å$^3$) | Atomic positions [27] | | | | | Ref. |
|---|---|---|---|---|---|---|---|---|---|
| | | | | Wyckoff | Element | x | y | z | |
| YCo$_2$Si$_2$ | 3.889 | 9.864 | 149.1863 | 2a | Y | 0 | 0 | 0 | This study |
| | | | | 4d | Co | 0 | 0.5 | 0.25 | |
| | | | | 4e | Si | 0 | 0 | 0.373 | |
| | 3.8875 | 9.734 | 147.1066 | | | | | | [27] |
| YNi$_2$Si$_2$ | 3.980 | 9.638 | 152.6698 | 2a | Y | 0 | 0 | 0 | This study |
| | | | | 4d | Co | 0 | 0.5 | 0.25 | |
| | | | | 4e | Si | 0 | 0 | 0.374 | |
| | 3.955 | 9.568 | 149.6629 | | | | | | [27] |
| YRu$_2$Si$_2$ | 4.177 | 9.708 | 169.3787 | 2a | Y | 0 | 0 | 0 | This study |
| | | | | 4d | Co | 0 | 0.5 | 0.25 | |
| | | | | 4e | Si | 0 | 0 | 0.368 | |
| | 4.158 | 9.546 | 165.0405 | | | | | | [27] |
| YRh$_2$Si$_2$ | 4.085 | 10.068 | 168.007 | 2a | Y | 0 | 0 | 0 | This study |
| | | | | 4d | Co | 0 | 0.5 | 0.25 | |
| | | | | 4e | Si | 0 | 0 | 0.378 | |
| | 4.018 | 9.897 | 159.7804 | | | | | | [27] |
| YPd$_2$Si$_2$ | 4.152 | 9.992 | 172.2531 | 2a | Y | 0 | 0 | 0 | This study |
| | | | | 4d | Co | 0 | 0.5 | 0.25 | |
| | | | | 4e | Si | 0 | 0 | 0.383 | |
| | 4.128 | 9.857 | 167.9671 | | | | | | [27] |
| YIr$_2$Si$_2$ | 4.081 | 10.046 | 167.3117 | 2a | Y | 0 | 0 | 0 | This study |
| | | | | 4d | Co | 0 | 0.5 | 0.25 | |
| | | | | 4e | Si | 0 | 0 | 0.364 | |
| | 4.048 | 9.815 | 160.8316 | | | | | | [27] |

### 3.2 Stiffness constants and mechanical insights

The elastic constants give a compressive understanding of how solid materials behave mechanically and dynamically. Several elementary solid-state phenomena such as nature of bonding existence between atoms, EOS and Phonon spectra are directly related to the elastic features [44]. These constants elaborate about a solid's stability, brittleness, anisotropy, ductility, and stiffness nature of materials [45]. Due to crystal symmetry a tetragonal crystal system contain six independent elastic constants namely; $C_{11}$, $C_{12}$, $C_{13}$, $C_{33}$, $C_{44}$ and $C_{66}$. The investigated results of the studied materials along with similar types of compounds are provided in Table 2. Elastic constants must meet Born criterion for tetragonal crystal structure [46].

$$\left. \begin{array}{l} C_{11} > 0,\ C_{44} > 0,\ C_{33} > 0,\ C_{66} > 0 \\ C_{11} + C_{33} - 2C_{13} > 0,\ C_{11} - C_{12} > 0 \\ 2(C_{11} + C_{12}) + C_{33} + 4C_{13} > 0 \end{array} \right\} \quad (1)$$

It is clear from **Table 2** that all the six materials YCo$_2$Si$_2$, YNi$_2$Si$_2$, YRu$_2$Si$_2$, YRh$_2$Si$_2$, YPd$_2$Si$_2$ and YIr$_2$Si$_2$ are mechanically stable in view of the fact that they fulfilled the above criteria.

**Table 2:** Calculated elastic constants ($C_{ij}$, in GPa) of superconducting YT$_2$Si$_2$ (T=Co, Ni, Ru, Rh, Pd, Ir) polymorphs along with similar types of compounds.

| Compounds | $C_{11}$ | $C_{12}$ | $C_{13}$ | $C_{33}$ | $C_{44}$ | $C_{66}$ | Ref. |
|---|---|---|---|---|---|---|---|
| YCo$_2$Si$_2$ | 265.59 | 85.51 | 97.45 | 238.82 | 113.55 | 117.97 | This study |
| YNi$_2$Si$_2$ | 239.88 | 54.81 | 94.01 | 248.98 | 97.36 | 56.81 | |
| YRu$_2$Si$_2$ | 266.88 | 97.56 | 93.87 | 222.26 | 78.87 | 114.27 | |
| YRh$_2$Si$_2$ | 256.03 | 124.66 | 105.12 | 257.14 | 95.59 | 122.91 | |
| YPd$_2$Si$_2$ | 215.70 | 67.01 | 118.47 | 234.04 | 75.98 | 47.79 | |
| YIr$_2$Si$_2$ | 333.20 | 140.95 | 102.28 | 302.25 | 112.00 | 153.64 | |
| | 313.00 | 130.00 | 98.00 | 305.00 | 99.00 | 14.00 | [19] |
| BaNi$_2$P$_2$ | 171.91 | 21.23 | 66.51 | 89.25 | 43.84 | 23.04 | [33] |
| BaIr$_2$P$_2$ | 241.36 | 106.68 | 47.12 | 119.82 | 30.05 | 124.97 | [31] |
| BaRh$_2$Ge$_2$ | 145.7 | 61.1 | 39.6 | 46.5 | 36.80 | 66.1 | [29] |
| SrRu$_2$P$_2$ | 222.22 | 90.34 | 47.30 | 78.38 | 35.40 | 121.26 | [30] |
| LaRu$_2$P$_2$ | 277.52 | 104.54 | 72.26 | 116.39 | 50.64 | 132.75 | [32] |
| LaPd$_2$Ge$_2$ | 178.36 | 49.78 | 89.53 | 157.90 | 54.46 | 32.29 | [35] |
| YRh$_2$Ge$_2$ | 217.0 | 86.9 | 98.4 | 201.0 | 47.5 | 74.9 | [29] |

Other basic parameters, called polycrystalline bulk moduli, like bulk modulus *B*, Shear modulus *G*, Young's modulus *E*, Pugh's ratio *B/G*, *B/C$_{44}$*, and H*v*, are very important for describing the mechanical features of materials, such as material's strength, ductility/brittleness, stiffness, ability to be machined, hardness, and so on. Table 3 contains of all of the bulk parameters, along with the value found in the previous investigation of similar 122 type's superconducting materials for comparative study [29-33]. Using the calculated elastic constants, the bulk modulus, shear modulus, Young's modulus, anisotropy, and hardness of superconducting YT$_2$Si$_2$ (T=Co, Ni, Ru, Rh, Pd, Ir) are determined by the Voigt-Reuss-Hill approximations [47]. Based on the Voigt-Reuss-Hill approximations, the bulk and shear modulus for tetragonal crystal system are as follows:

$$B_V = \frac{2C_{11} + 2C_{12} + C_{33} + 4C_{13}}{9} \quad (2)$$

$$B_R = \frac{C^2}{M} \quad (3)$$

$$G_V = \frac{M + 3C_{11} - 3C_{12} + 12C_{44} + 6C_{66}}{30} \quad (4)$$

$$G_R = \frac{15}{\left[\frac{18B_V}{C^2} + \frac{6}{(C_{11} - C_{12})} + \frac{6}{C_{44}} + \frac{3}{C_{66}}\right]} \quad (5)$$

Where, $M = C_{11} + C_{12} + 2C_{33} - 4C_{13}$ and $C^2 = (C_{11} + C_{12})C_{33} - 2C_{13}^2$.

The average value of *B* and *G* given taken by Hill is giving by the following two relations,

$$B = \frac{1}{2}(B_R + B_v) \quad (6)$$

$$G = \frac{1}{2}(G_v + G_R) \qquad (7)$$

Now we can calculate the Young's modulus ($E$), and Poisson's ratio ($v$) from the relations below,

$$E = \frac{9GB}{3B + G} \qquad (8)$$

$$v = \frac{3B - 2G}{2(3B + G)} \qquad (9)$$

The anisotropic behaviors of a crystal can be estimated by the following equation [48],

$$A^U = \frac{5G_V}{G_R} + \frac{B_V}{B_R} - 6 \qquad (10)$$

The hardness of materials can be predicted by using the following relation [49],

$$H_V = 2(K^2 G)^{0.585} - 3 \qquad (11)$$

**Table 3:** Calculated polycrystalline elastic constants ($C_{ij}$, in GPa) and hardness, $H_v$ (GPa) of superconducting YT$_2$Si$_2$ (T=Co, Ni, Ru, Rh, Pd, Ir) polymorphs along with similar types of compounds.

| Phases | $B_V$ | $B_R$ | $B$ | $G_V$ | $G_R$ | $G$ | $E$ | $H_v$ | Ref. |
|---|---|---|---|---|---|---|---|---|---|
| YCo$_2$Si$_2$ | 147.87 | 147.76 | 147.81 | 101.65 | 98.28 | 99.97 | 244.74 | 15.72 | This study |
| YNi$_2$Si$_2$ | 134.93 | 133.69 | 134.31 | 82.70 | 78.76 | 80.78 | 201.88 | 11.40 | |
| YRu$_2$Si$_2$ | 147.41 | 146.21 | 146.81 | 85.78 | 83.50 | 84.64 | 212.99 | 11.09 | |
| YRh$_2$Si$_2$ | 159.88 | 159.72 | 159.81 | 91.77 | 87.13 | 89.79 | 226.88 | 11.15 | |
| YPd$_2$Si$_2$ | 141.48 | 137.57 | 139.53 | 64.05 | 60.26 | 62.16 | 162.37 | 5.70 | |
| YIr$_2$Si$_2$ | 184.41 | 182.80 | 183.61 | 117.07 | 114.13 | 115.60 | 286.64 | 15.74 | |
| | 175.90 | 175.40 | 175.60 | 82.70 | 45.30 | 64.00 | 171.20 | 3.99 | [19] |
| BaNi$_2$P$_2$ | - | - | 80.92 | - | - | 35.87 | 93.77 | 3.27 | [33] |
| BaIr$_2$P$_2$ | - | - | 102.46 | - | - | 55.04 | 140.04 | 6.92 | [31] |
| BaRh$_2$Ge$_2$ | 68.72 | 45.83 | 57.30 | 41.11 | 31.89 | 36.50 | 90.3 | 3.56 | [29] |
| SrRu$_2$P$_2$ | 99.19 | 71.48 | 85.33 | 60.94 | 45.87 | 53.40 | 132.55 | 8.83 | [30] |
| LaRu$_2$P$_2$ | - | - | 117.19 | - | - | 68.15 | 171.25 | 9.53 | [32] |
| LaPd$_2$Ge$_2$ | - | - | 107.80 | - | - | 45.26 | 119.11 | 3.74 | [35] |
| YRh$_2$Ge$_2$ | 133.60 | 133.59 | 133.60 | 57.40 | 55.53 | 56.50 | 148.6 | 7.09 | [29] |

The fundamental bulk characteristic $B$, defined as volume stress to volume strain, reveals a material's ability to resist plastic deformation. A high value for $B$ indicates that the material is capable of resisting significant amounts of plastic or volume deformation. From Table 3 and Fig. 2(b) we have observed that among that all the compounds, YIr$_2$Si$_2$ has a highest value of $B$, which ensured that this phase is more capable to resist plastic deformation. The sequence of bulk modulus of all the compounds listed in Table 3 is YIr$_2$Si$_2$>YRh$_2$Si$_2$> YCo$_2$Si$_2$>YRu$_2$Si$_2$>YNi$_2$Si$_2$>YPd$_2$Si$_2$> YRh$_2$Ge$_2$> LaRu$_2$P$_2$> LaPd$_2$Ge$_2$> BaIr$_2$P$_2$> SrRu$_2$P$_2$> BaNi$_2$P$_2$> BaRh$_2$Ge$_2$. From this analysis we have seen that the studied compounds YT$_2$Si$_2$ (T=Co, Ni, Ru, Rh, Pd, Ir) have more ability to resist plastic deformation than the previous studied materials [29-35].

The shear modulus, $G$, is another essential bulk factor that determines the material's resistance to shear deformation as the ratio of shear stress to shear strain. From Table 3 and Fig. 2(b), we can see that $YIr_2Si_2$ has a higher $G$ than the other compounds listed in Table 3, which means it can resist shear deformation better than the others. The sequence of shear modulus of the studied compounds is $YIr_2Si_2 > YCo_2Si_2 > YRh_2Si_2 > YRu_2Si_2 > YNi_2Si_2 > YPd_2Si_2$.

Young's modulus, $E$, is a crucial bulk quantity that assesses a material's resistance to longitudinal stress. In addition, it has managed the thermal shock resistance of a periodic crystal. The system where $E$ is inversely proportional to $R$ is referred to as the crucial thermal shock coefficient [50], which indicates that the lower the value of $E$, the greater the resistance to thermal shock. Here the lower value of $YPd_2Si_2$ ensured that it has high ability to resist the thermal shock than the other compounds studied (Table 3 and Fig. 2(b)) in this work. Comparing among all the compounds as shown in Fig. 2(b) and Table 3, we can see that the phase $YIr_2Si_2$ has highest $E$ value indicating that this phase has highest ability to resist longitudinal stress than all the rest ones. On the other hand, $BaRh_2Ge_2$ has the smallest value of $E$ among all of the phases (see Fig. 2(b)), which indicates that $BaRh_2Ge_2$ is more suitable to be employed as a thermal barrier coating (TBC) material compared to others. The Young's modulus sequence of all the compounds is $YIr_2Si_2 > YCo_2Si_2 > YRh_2Si_2 > YRu_2Si_2 > YNi_2Si_2 > YPd_2Si_2 > LaRu_2P_2 > YRh_2Ge_2 > BaIr_2P_2 > SrRu_2P_2 > LaPd_2Ge_2 > BaNi_2P_2 > BaRh_2Ge_2$.

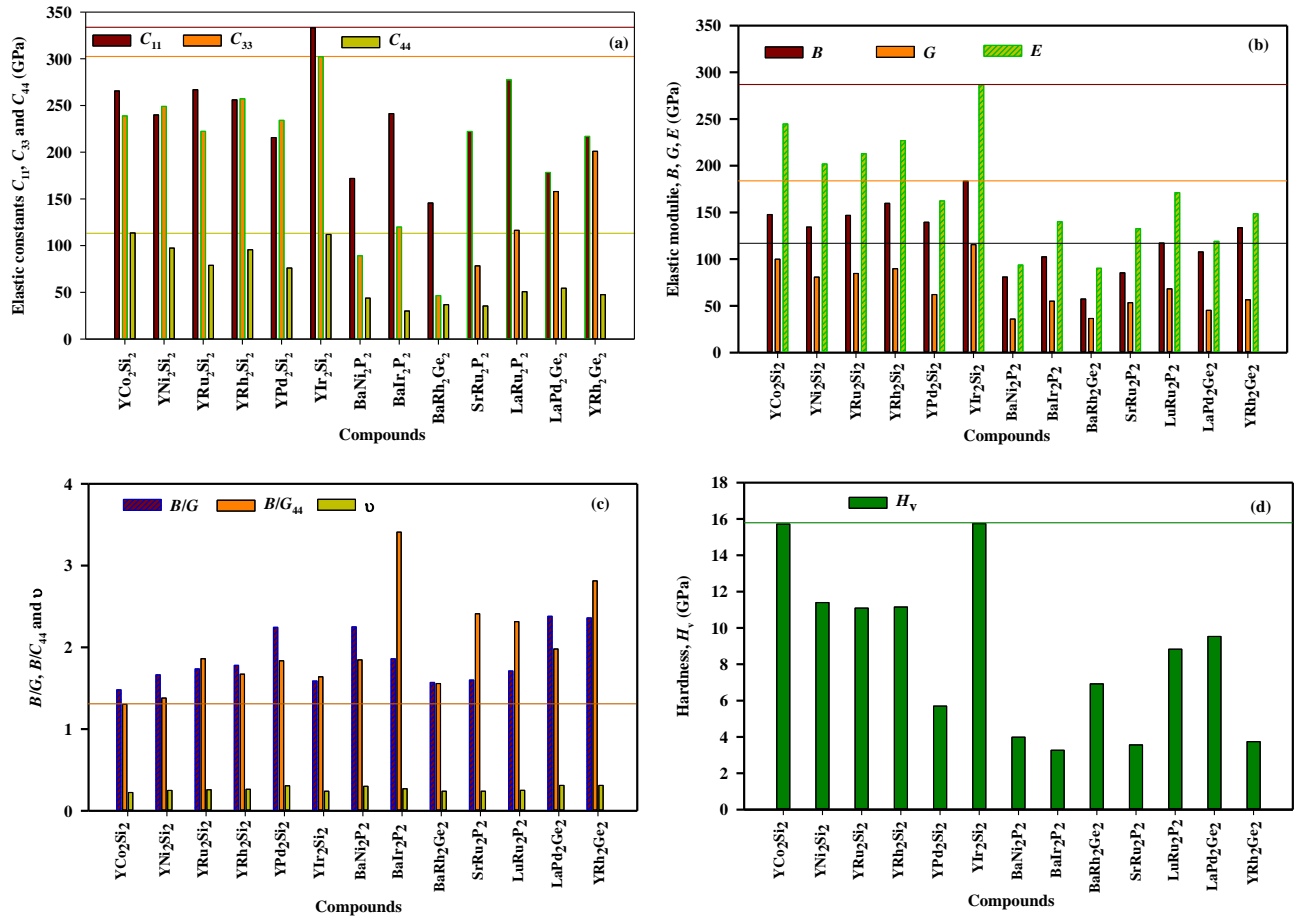

**Fig. 2:** Bulk features of superconducting $YT_2Si_2$ (T=Co, Ni, Ru, Rh, Pd, Ir).

**Table 4:** Calculated Pugh's ratio ($B/G$), Poisson's ratio ($\nu$), machinability index ($\mu m = B/C_{44}$), elastic anisotropic factors ($A$ and $A^U$) of superconducting $YT_2Si_2$ (T=Co, Ni, Ru, Rh, Pd, Ir) polymorphs with some 122 type superconductors.

| Phases | Status | B/G | ν | B/C$_{44}$ | A | A$^U$ | Ref. |
|---|---|---|---|---|---|---|---|
| YCo$_2$Si$_2$ | Brittle | 1.479 | 0.224 | 1.302 | 1.261 | 0.173 | This study |
| YNi$_2$Si$_2$ | Brittle | 1.663 | 0.249 | 1.379 | 1.052 | 0.259 | |
| YRu$_2$Si$_2$ | Brittle | 1.735 | 0.258 | 1.861 | 0.932 | 0.145 | |
| YRh$_2$Si$_2$ | Ductile | 1.780 | 0.263 | 1.672 | 1.455 | 0.227 | |
| YPd$_2$Si$_2$ | Ductile | 2.245 | 0.306 | 1.836 | 1.022 | 0.343 | |
| YIr$_2$Si$_2$ | Brittle | 1.588 | 0.239 | 1.639 | 1.269 | 0.138 | |
| BaNi$_2$P$_2$ | Ductile | 2.25 | 0.30 | 1.846 | 0.582 | 1.600 | [33] |
| BaIr$_2$P$_2$ | Ductile | 1.86 | 0.27 | 3.409 | 0.446 | 2.090 | [31] |
| BaRh$_2$Ge$_2$ | Brittle | 1.57 | 0.24 | 1.557 | 0.869 | 1.945 | [29] |
| SrRu$_2$P$_2$ | Brittle | 1.60 | 0.24 | 2.410 | 0.537 | 2.020 | [30] |
| LaRu$_2$P$_2$ | Brittle | 1.71 | 0.25 | 2.314 | 0.586 | 1.350 | [32] |
| LaPd$_2$Ge$_2$ | Ductile | 2.38 | 0.31 | 1.979 | 0.847 | 0.480 | [35] |
| YRh$_2$Ge$_2$ | Ductile | 2.36 | 0.31 | 2.813 | 0.730 | 0.168 | [29] |

In order to predict whether a material would be ductile or brittle, scientist Pugh used a formula between bulk modulus and shear modulus (B/G): if B/G is greater than 1.75, the material in question would be ductile, otherwise it would be brittle [8]. Another important bulk parameter known as Poisson ratio, ν predict the ductile/brittle nature of material. According to this rule a material will show ductile/brittle nature when the value of ν will be ν> 0.26 and ν<0.26 respectively [9]. According to above criteria the studied compounds YCo$_2$Si$_2$, YNi$_2$Si$_2$, and YIr$_2$Si$_2$ show brittle nature and YRh$_2$Si$_2$ and YPd$_2$Si$_2$ show ductile nature (see **Table 3** and **Fig. 3**) where YRu$_2$Si$_2$ lies on the brittle/ductile border line.

The machinability of a material is a key consideration in today's industrial sector, where the goal is to maximize output while minimizing expenses. The machinability index is influenced by a wide variety of factors, including the cutting form, the type of cutting, the rigidity, the hardness, and the capacity of the machine tool [10]. This useful parameter is denoted by μ and defined [11] by the ratio of bulk modulus, B to shear resistance, $C_{44}$ i.e

$$\mu = B/C_{44} \qquad (13)$$

This index also measures the material's lubricating properties and its plasticity [12-13]. High $B/C_{44}$ means that a material has great lubricating properties, little friction, and a high plastic strain. From **Fig. 2 (c)** it is very clear that among the six materials YRu$_2$Si$_2$ has high machinability index indicating that it is very good machinable than all the others materials. The findings of this study are intriguing because they show that materials with low bulk, Young's, and shear moduli still have very high machinability and can be effectively used in industry.

Elastic anisotropy factor, A, also known as the Zener anisotropy factor, defines the bulk properties of a crystal system that depend on direction. If the properties of a crystal are uniform in all directions, then the crystal is isotropic; if the attributes vary in different directions, the crystal is anisotropic. The factor $A=1$ indicates the isotropic nature of a crystal, in other cases the crystal will be anisotropic. From our calculations we can say that all the compounds show anisotropic nature according to above criteria, where $YRh_2Si_2$ exhibits the largest anisotropic nature among the studied compounds.

Material hardness is another macroscopic bulk parameter that has a lot of uses in business and technology. A material with a high hardness has a greater capacity to withstand plastic deformation than one with a low hardness which is appropriate for plastic deformation and it becomes more machinable. A material will be more machinable and damage-tolerant when the value of hardness lies between 2 to 8 [14]. Analyzing **Fig. 2(d)** we can say that the material $YIr_2Si_2$ highest hardness compared to other five compounds under study indicating that it has highest ability to resist plastic deformation than rest compounds. Consequently the low hardness of $YPd_2Si_2$ among the six components and among the ten compounds $BaIr_2P_2$ confirming that it is more machinable as shown in **Fig. 2 (d).**

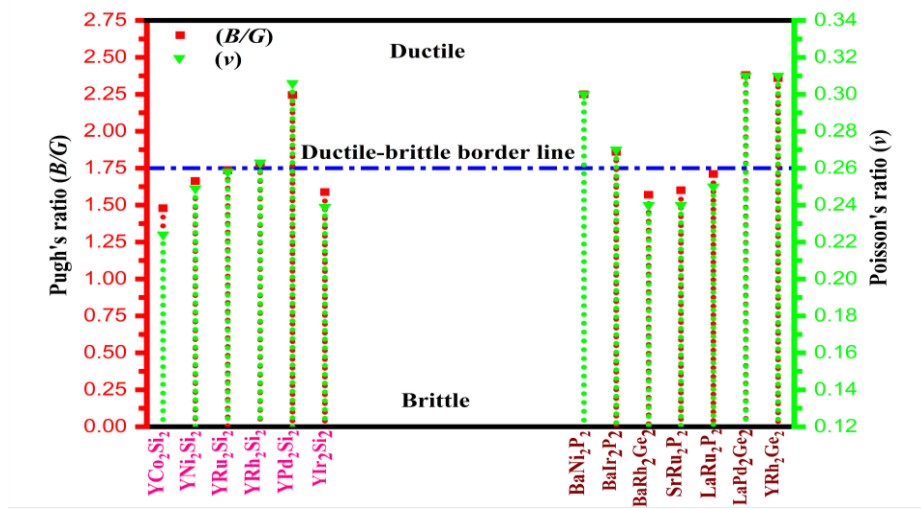

**Fig. 3:** Ductile/brittle nature of superconducting $YT_2Si_2$ (T=Co, Ni, Ru, Rh, Pd, Ir).

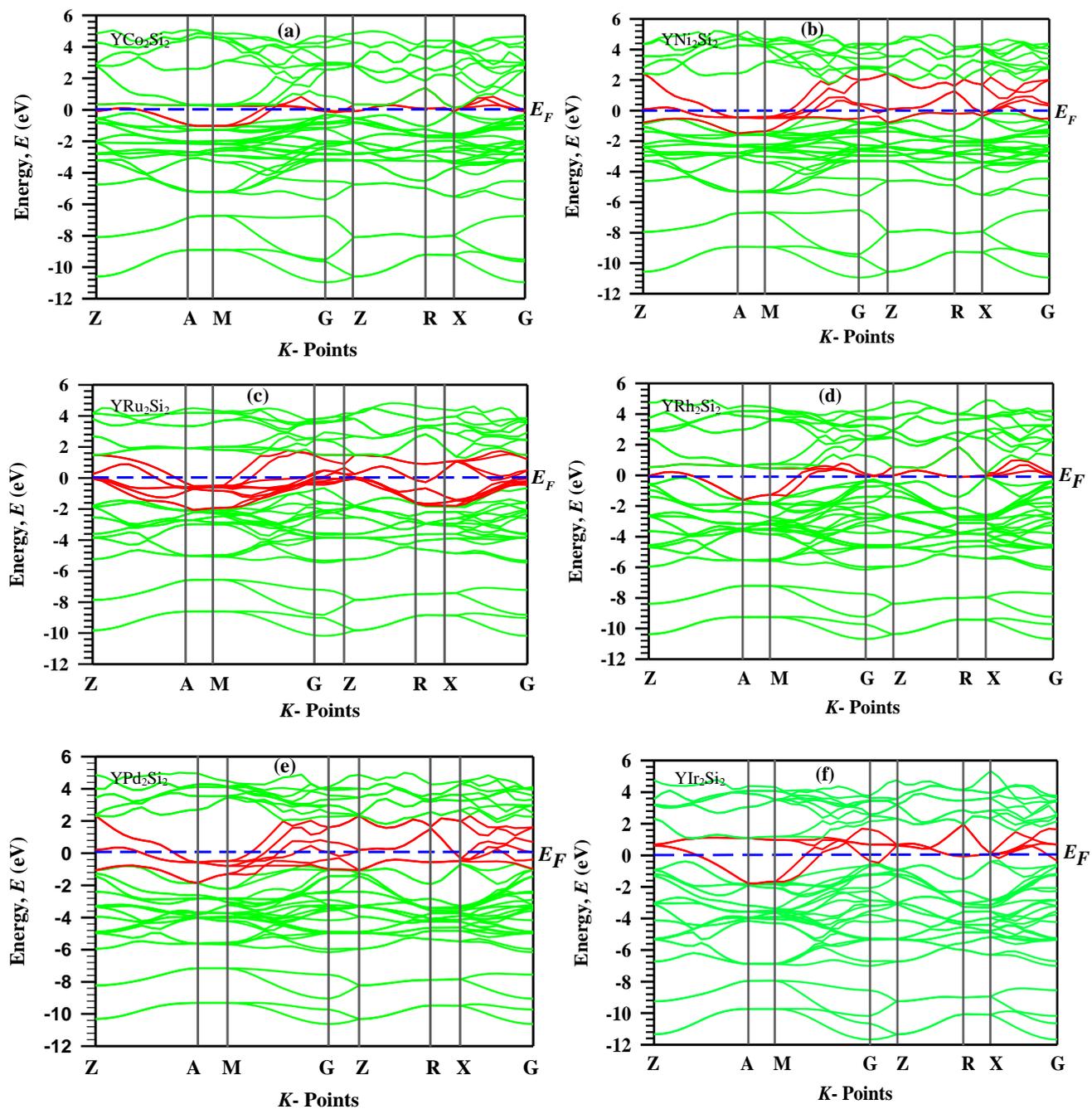

**Fig. 4:** Electronic Band structures of superconducting silicides YT$_2$Si$_2$ (T=Co, Ni, Ru, Rh, Pd, Ir).

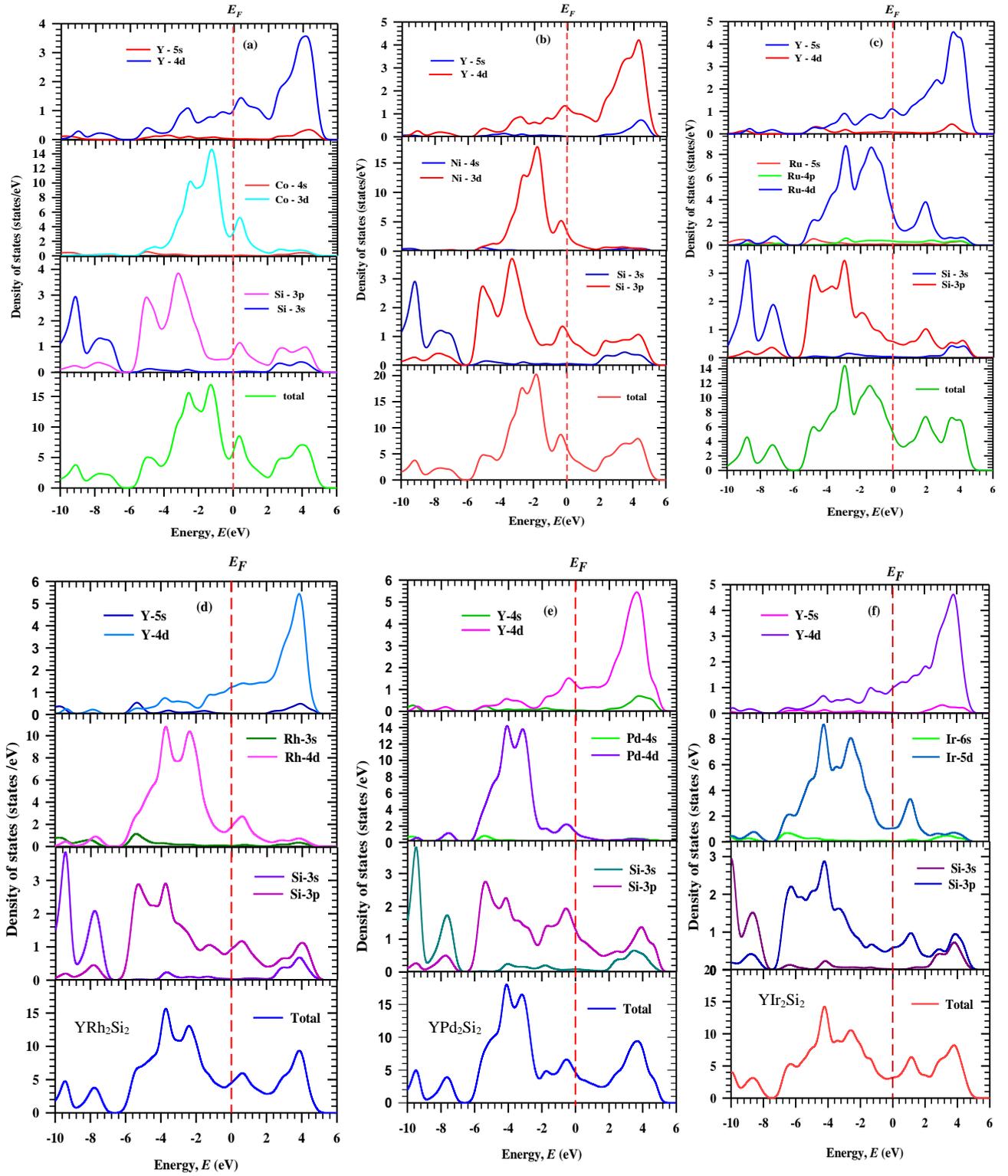

**Fig. 5:** Total and partial density of states of superconducting silicides YT$_2$Si$_2$ (T=Co, Ni, Ru, Rh, Pd, Ir).

## 5: Optical Properties

To gain insight into the electronic structure of the material the optical properties is one of the most important techniques. We have discussed the optical properties of YT$_2$Si$_2$ (X=Co, Ni, Ru, Rh, Pd, Ir) compounds in this section. To predict the optical function of YT$_2$Si$_2$ (X=Co, Ni, Ru, Rh, Pd, Ir) with various phonon energies [79] we have used the frequency-dependent dielectric function $\varepsilon(\omega) = \varepsilon_1(\omega) + i\varepsilon_2(\omega)$, which is closely related to the electronic configuration where the real $\varepsilon_1(\omega)$ and imaginary $\varepsilon_2(\omega)$ dielectric functions are correlated by the Kramers-Kronig relation. The real part $\varepsilon_1(\omega)$ can be calculated by the equation [80],

$$\varepsilon_1(\omega) = 1 + \frac{2}{\pi} P \int_0^\infty \frac{\omega' \varepsilon_2(\omega')}{\omega'^2 - \omega^2} d\omega' \tag{14}$$

Where $\omega$ is for the light frequency and $P$ for the principle value of the integral part. The real $\varepsilon_1(\omega)$ and imaginary $\varepsilon_2(\omega)$ dielectric functions is obtained from the momentum matrix elements and taking into account every possible transition between the occupied and unoccupied electronic state, the imaginary part $\varepsilon_2(\omega)$ is calculated directly using [81]-

$$\varepsilon_2(\omega) = \frac{2e^2\pi}{\Omega\varepsilon_0} \sum_{k,v,c} |\psi_k^c|u.r|\psi_k^v|^2 \delta(E_k^c - E_k^v - E) \tag{15}$$

Where $\psi_k^c$ and $\psi_k^v$ are the wave function at k of conduction and valance band, u denotes a vector defining the polarization of the incident electric field, e is denoted as the electric charge and $\omega$ denotes the light frequency respectively. The optical functions of YT$_2$Si$_2$ (X=Co, Ni, Ru, Rh, Pd, Ir) are estimated for phonon energies up to 20 eV which are shown in **Fig.11.** We have used 0.5 eV Gaussian smearing for all calculations because it spreads out of the Fermi level so that k-points are more efficient on the Fermi surface.

**Fig.**10(a) and 10(b) show the expected real and imaginary parts of dielectric functions for all phases respectively. The large negative value of $\varepsilon_1(\omega)$ for any of these phases implies Drude-like metallic behavior. From above the imaginary portion $\varepsilon_2(\omega)$ approaches to zero, while the real part $\varepsilon_1(\omega)$ reaches through zero from below also show metallic nature. The plasma frequency 4 eV and damping (relaxation) energy 0.05 eV have been used. The main characteristics of our measured optical spectra of YT$_2$Si$_2$ (X=Co, Ni, Ru, Rh, Pd, Ir) are remarkably comparable, despite some variation in peak heights and locations.

The refractive index, which has no dimensions, quantifies and which is defined as how much light is bent or refracted as it enters a substance [88]. The idea of the refractive index of an optical material is significant for its use in optical tools such as photonic crystals, waveguides etc. **Fig.** 10(c) and 10(d) display the refractive indices (real and imaginary portion) of the YT$_2$Si$_2$ (X=Co, Ni, Ru, Rh, Pd, Ir) compounds. The similar contribution found for all compounds in the whole graph. Both the phases process the high refractive index in the infrared and visible area and ultraviolet regions these values are gradually decreased.

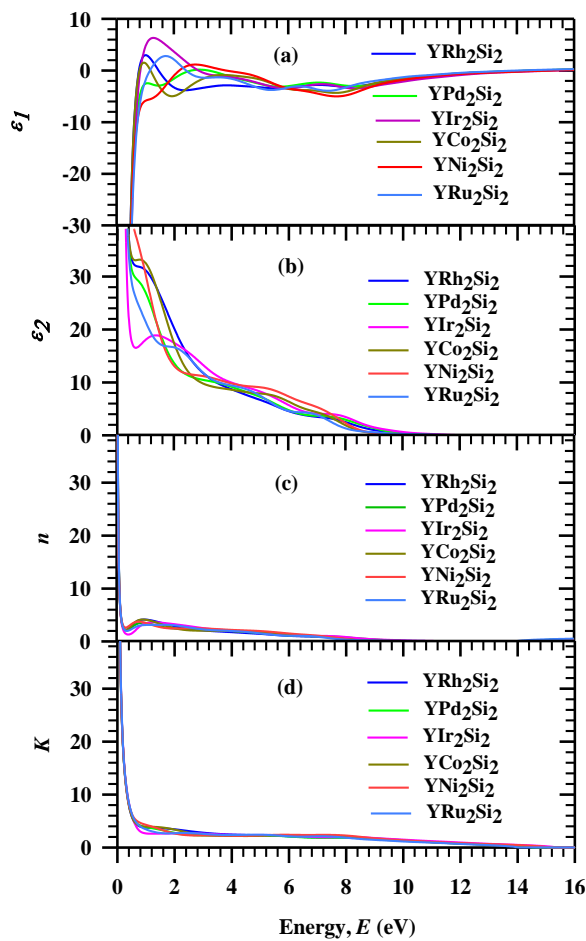

**Fig. 10:** (a and b) Dielectric functions, (c) refractive indices and (d) extension coefficient of $YT_2Si_2$ (T=Co, Ni, Ru, Rh, Pd, Ir).

The absorption coefficient is a necessary factor that indicates how much light of specific energy (wavelength) can enter a material before being absorbed, and it provides information on solar energy conversion efficiency. In **Fig.11(a)** absorption spectra for $YT_2Si_2$ (X=Co, Ni, Ru, Rh, Pd, Ir) are shown and they all start at 0 eV because their metallic nature,. According to **Fig. 11(a)**, for all of these compounds, the nature of the absorption curve is same, and many absorption peaks were found at various energy levels. However, all of these compounds possess good absorption coefficient.

Conductivity of a material refers to the ability to transport electron with the variation of photon energy [82]. It helps us to identify the material will be conductor, semiconductor or superconductor. The investigated conductivity spectra of $YT_2Si_2$ (X=Co, Ni, Ru, Rh, Pd, Ir) with photon energy are shown in **Fig.11(b)**. From this figure it has been seen that photoconductivity starts with zero photon energy due to the reason that the materials have no band gap which is evident from the band structure indicating the metallic behaviors of these phases at energy 7.5eV the conductivity shows higher value; after that it gradually decreases and after energy about 13eV the conductivity vanishes for all compounds.

Reflectivity is the measure of light reflected by the materials upon incident on it. It is obtained by the ratio of the energy of the wave reflected from a surface to the energy of the wave incident on the surface [83]. The reflectivity of $YT_2Si_2$ (X=Co, Ni, Ru, Rh, Pd, Ir) as a function of photon energy is illustrated in Fig.11(c). It has been seen that reflectivity starts from 0.68 for $YNi_2Si_2$ and all other phases have the roughly similar spectra and the reflectivity starts between(5.2-6.2) eV with zero photon energy. The observed value represents good reflectivity in infrared and visible region. In ultraviolet region the absorption coefficient gradually decreases.

The energy loss function is a key factor in the dielectric formalism, which is utilized to clarify the optical dielectric formalism, which is utilized to clarify the optical created by fast charges in solid material. It is defined as the energy loss by a fast electron when it traverses in the material [84]. The frequency at which maximum energy loss happened is known as the bulk plasma frequency $\omega_p$ of the material which emerges at $\varepsilon_2(\omega)<1$ and $\varepsilon_1(\omega)= 0$ [85–87]. The energy loss spectra show that all phases have the roughly similar plasma frequency $\omega_p$ with the approximated value (13-15.2)eV respectively indicating that all phases will be transparent when the incident photon energies are greater than (13-15.2) eV.

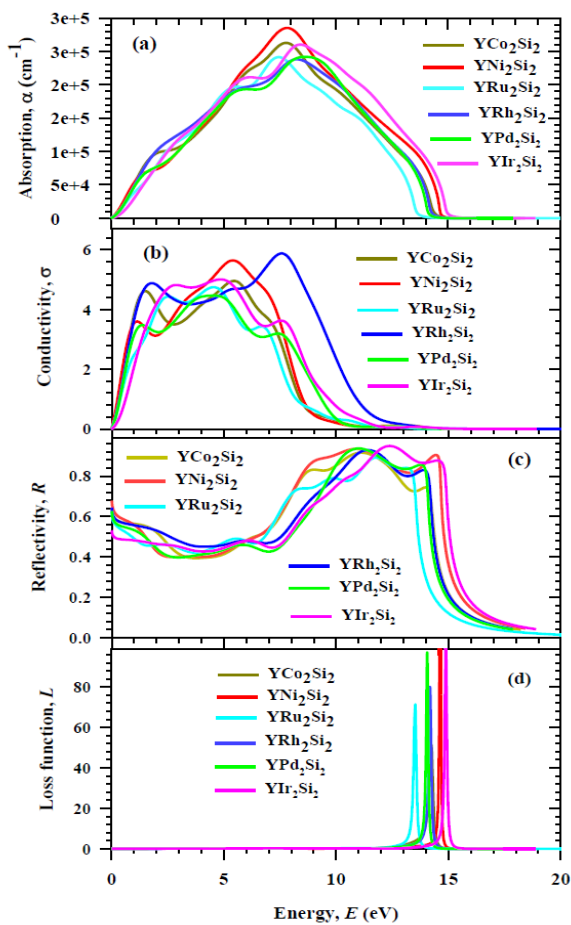

**Fig. 11:** (a) Absorption, (b) conductivity, (c) reflectivity and (d) loss function of $YT_2Si_2$ (T=Co, Ni, Ru, Rh, Pd, Ir).

## 6. Thermodynamic Properties

The several solid-state phenomena of the crystalline materials are directly described by very essential parameters called thermodynamic properties. These features included the Debye temperature, minimum thermal conductivity, melting temperature and Dulong-Petit limit. The Debye temperature is usually associated to the crystal's highest normal mode of vibration and provides deep insight into several vital thermodynamic features included thermal expansion, specific heat, melting point etc. Several approximations have been applied to determine the Debye temperature of materials. Here we have utilized the single elastic stiffness constant values to explore the Debye temperature of YT$_2$Si$_2$ (T=Co, Ni, Ru, Rh, Pd, Ir). The calculated values of Debye temperature $\theta_D$, minimum thermal conductivity and melting temperature of YT$_2$Si$_2$ (T=Co, Ni, Ru, Rh, Pd, Ir) with some 122 type superconductors are listed in Table 7.

In a particular temperature or minimum temperature, Debye temperature is known as the maximum frequency mode of vibration. The Debye temperature ($\theta_D$) is not an accurately calculated parameter and for this reason to evaluate the Debye temperature of a material the various data can be used but we have used to elastic modulus data to calculate the Debye temperature. By using the following equations, the Debye temperature has been estimated [86-90],

$$\theta_D = \frac{h}{k_B} \left[ \left(\frac{3n}{4\pi}\right) \frac{N_A \rho}{M} \right]^{1/3} \times v_m \qquad (16)$$

In this expression $h$ and $K_B$ indicate the Planck and Boltzman constant, $N_A$ indicates the Avogrado's number, $\rho$ indicates the density, M indicates as the Molecular weight and $n$ indicates the number of atoms contained in the unit cell of (T=Co, Ni, Ru, Rh, Pd, Ir).

$$v_m = \left[ \frac{1}{3} \left( \frac{2}{v_t^3} + \frac{1}{v_l^3} \right) \right]^{-1/3} \qquad (17)$$

Furthermore, $v_m$ be the average elastic (sound) wave velocity which is used in the equation of Debye temperature calculation. $v_m$ can be determined by using the value of $v_t$ and $v_l$ and also known as the transverse and longitudinal sound velocities.

$$v_l = \left(\frac{3B+4G}{3\rho}\right)^{1/2} \qquad (18)$$

$$v_t = \left(\frac{G}{\rho}\right)^{1/2} \qquad (19)$$

The minimum thermal conductivity ($K_{min}$) is an essential property of solid which reveals the conduction of heat in a material. Thermal conductivity decreases for increasing the temperature and reaches at a certain point called minimum thermal conductivity [91]. The minimum thermal conductivity can be obtained by the following equation [92]:

$$K_{min} = k_B v_m \left(\frac{nN_A\rho}{M}\right)^{2/3} \qquad (20)$$

Where, N is the number of atoms per unit cell and $N_A$ is the Avogadro number. The calculated minimum thermal conductivity is listed in Table 6. For the low value of Debye temperature the minimum thermal conductivity indicates the low value according to the thumb rule [93]. The melting temperature becomes one of the important properties of a solid crystal. And the solid alters its state to liquid at atmospheric pressure and this fact is well-known as melting point of solid. At melting temperature the solid and liquid also found in equilibrium state. Fine et al. suggested the following empirical equation for cubic substance, which is applied to compute the melting temperature: [94],

$T_m = 553 + 5.91 C_{11}$ (21)

**Table 5:** The evaluated density $\rho$, transverse sound velocity $v_t$, longitudinal sound velocity $v_l$, average sound velocity $v_m$ and Debye temperature $\theta_D$, the minimum thermal conductivity and melting temperature of $YT_2Si_2$ (T=Co, Ni, Ru, Rh, Pd, Ir) with some 122 type superconductors.

| Compounds | $\rho$ (kg/m³) | $v_t$ (m/s) | $v_l$ (m/s) | $v_m$ (m/s) | $\theta_D$ (K) | $K_{min}$ (Wm⁻¹K⁻¹) | $T_m$ (K) | Ref. |
|---|---|---|---|---|---|---|---|---|
| YCo$_2$Si$_2$ | 5.835 | 4132.81 | 6930.17 | 4574.59 | 553.66 | 1.04 | 1709.9 | This work |
| YNi$_2$Si$_2$ | 5.709 | 3761.59 | 6510.92 | 4179.16 | 501.94 | 0.94 | 1647.11 | |
| YRu$_2$Si$_2$ | 6.807 | 3526.22 | 6176.29 | 3918.56 | 454.44 | 0.87 | 1688.03 | |
| YRh$_2$Si$_2$ | 6.936 | 3597.99 | 6348.33 | 4000.78 | 465.33 | 0.84 | 1707.80 | |
| YPd$_2$Si$_2$ | 6.899 | 3001.67 | 5677.85 | 3355.38 | 387.01 | 0.69 | 1552.16 | |
| YIr$_2$Si$_2$ | 10.509 | 3316.67 | 5670.70 | 3677.84 | 428.18 | 0.78 | 2006.98 | |
| BaNi$_2$P$_2$ | 5.68 | 2513.23 | 4761.23 | 2876.93 | 323.70 | 0.56 | 1003.60 | [33] |
| BaIr$_2$P$_2$ | 9.60 | 2394.43 | 4279.87 | 2669.30 | 293.06 | 0.49 | 1257.81 | [31] |
| BaRh$_2$Ge$_2$ | 7.42 | 2217.91 | 3779.05 | 2458.68 | 262.07 | 0.43 | 860.85 | [29] |
| SrRh$_2$Ge$_2$ | 7.49 | 2692.52 | 4607.41 | 2985.98 | 330.98 | 0.57 | 1047.90 | |
| SrRu$_2$P$_2$ | 6.27 | 2918.34 | 4996.48 | 3236.66 | 363.86 | 0.63 | 1138.23 | [30] |
| LaPd$_2$Ge$_2$ | 8.09 | 2365.28 | 4559.00 | 2647.33 | 291.69 | 0.50 | 1125.93 | [35] |
| YRh$_2$Ge$_2$ | 8.04 | 2650.91 | 5097.71 | 2966.57 | 336.38 | 0.59 | 1306.50 | [29] |

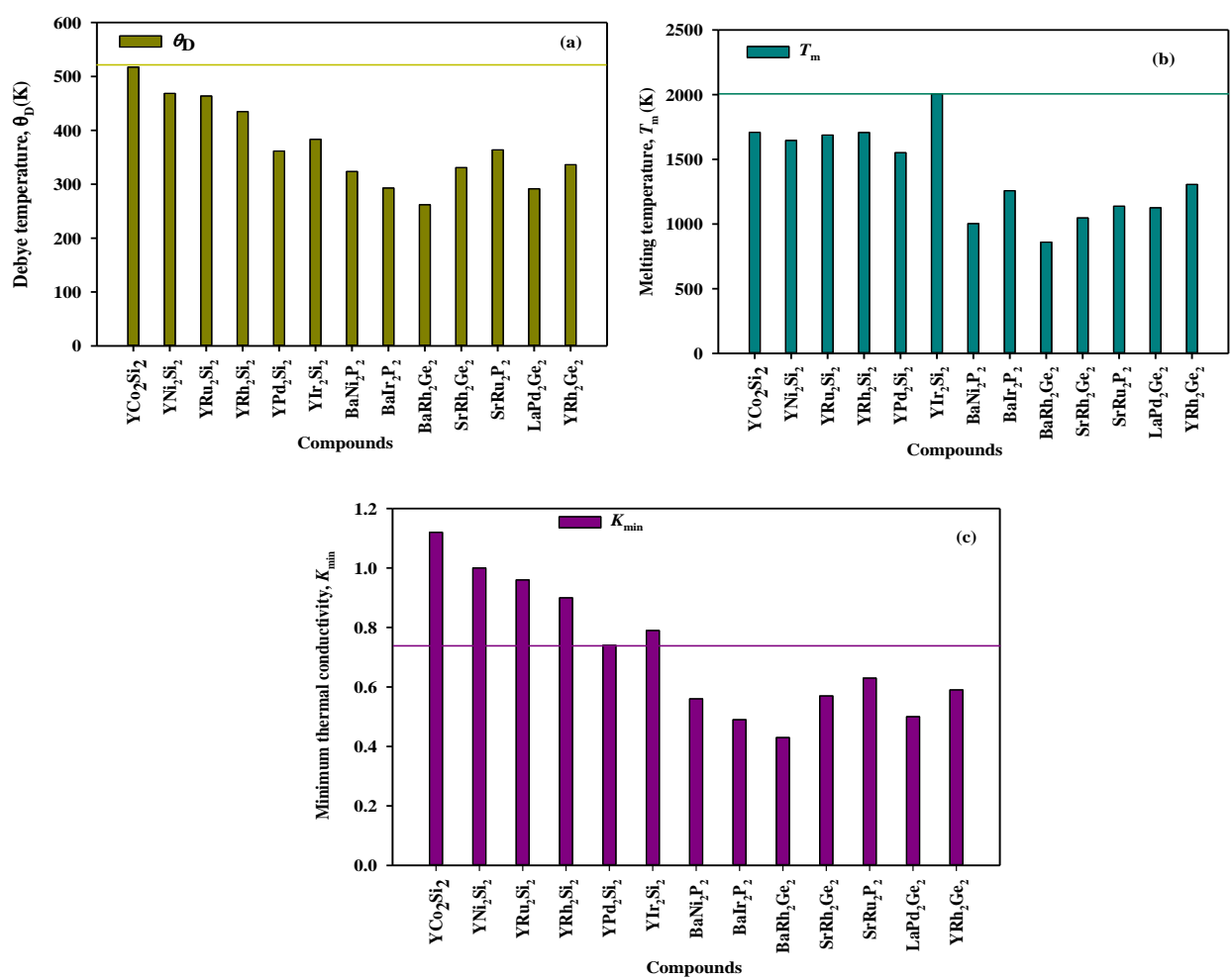

**Fig.12:** Thermodynamic properties of YT$_2$Si$_2$ (T=Co, Ni, Ru, Rh, Pd, Ir).